\documentstyle[11pt,amssymb]{article}

\makeatletter
\@addtoreset{equation}{section}
\makeatother

\def\dalemb#1#2{{\vbox{\hrule height .#2pt
        \hbox{\vrule width.#2pt height#1pt \kern#1pt
                \vrule width.#2pt}
        \hrule height.#2pt}}}
\def\square{\mathord{\dalemb{6.8}{7}\hbox{\hskip1pt}}}

\def\cF{{\cal F}}
\def\cA{{\cal A}}

\def\0{{\sst{(0)}}}
\def\1{{\sst{(1)}}}
\def\2{{\sst{(2)}}}
\def\3{{\sst{(3)}}}
\def\4{{\sst{(4)}}}
\def\5{{\sst{(5)}}}
\def\6{{\sst{(6)}}}
\def\7{{\sst{(7)}}}
\def\8{{\sst{(8)}}}

\def\ep{\epsilon}
\def\td{\tilde}
\def\wtd{\widetilde}

\let\a=\alpha \let\b=\beta

\def\nn{\nonumber} \def\bd{\begin{document}} \def\ed{\end{document}}
\def\ds{\documentstyle} \let\fr=\frac \let\bl=\bigl \let\br=\bigr
\let\Br=\Bigr \let\Bl=\Bigl 
\let\bm=\bibitem
\let\na=\nabla
\let\pa=\partial \let\ov=\overline 
\newcommand{\be}{\begin{equation}} 
\newcommand{\ee}{\end{equation}} 
\def\ba{\begin{array}}
\def\ea{\end{array}}
\def\ft#1#2{{\textstyle{{\scriptstyle #1}\over {\scriptstyle #2}}}}
\def\fft#1#2{{#1 \over #2}}
\def\del{\partial}
\def\sst#1{{\scriptscriptstyle #1}}
\def\oneone{\rlap 1\mkern4mu{\rm l}}
\def\ie{{\it i.e.\ }}
\def\via{{\it via}}
\def\semi{{\ltimes}}
\def\str{{\rm str}}
\def\jm{{\rm j}}
\def\im{{\rm i}}
\def\mapright#1{\smash{\mathop{-\!\!\!-\!\!\!-\!\!\!-\!\!\!-\!\!\!
             \longrightarrow}\limits^{#1}}}
\def\maprightt#1#2{\smash{\mathop{-\!\!\!-\!\!\!-\!\!\!-\!\!\!-\!\!\!
             \longrightarrow}\limits^{#1}_{#2}}}

\newcommand{\ho}[1]{$\, ^{#1}$}
\newcommand{\hoch}[1]{$\, ^{#1}$}
\newcommand{\bea}{\begin{eqnarray}} 
\newcommand{\eea}{\end{eqnarray}} 
\newcommand{\ra}{\rightarrow}
\newcommand{\lra}{\longrightarrow}
\newcommand{\Lra}{\Leftrightarrow}
\newcommand{\ap}{\alpha^\prime}
\newcommand{\bp}{\tilde \beta^\prime}
\newcommand{\tr}{{\rm tr} }
\newcommand{\Tr}{{\rm Tr} } 
\newcommand{\NP}{Nucl. Phys. }
\newcommand{\tamphys}{\it Center for Theoretical Physics\\
Texas A\&M University, College Station, Texas 77843}
\newcommand{\ens}{\it Laboratoire de Physique Th\'eorique de l'\'Ecole
Normale Sup\'erieure\hoch{2,3}\\
24 Rue Lhomond - 75231 Paris CEDEX 05}
\newcommand{\upenn}{\it Department of Physics and Astronomy\\
University of Pennsylvania, Philadelphia, Pennsylvania 19104}

\newcommand{\auth}{E. Lima\hoch{\dagger}, H. L\"u\hoch{\dagger1},
B.A. Ovrut\hoch{\dagger2} and C.N. Pope\hoch{\ddagger3}}

\begin{document}
\begin{flushright}
\hfill{CTP TAMU-08/98}\\
\hfill{UPR-834-T}\\
\hfill{hep-th/9903001}\\
\hfill{March 1999}\\
\end{flushright}


\begin{center}
{ \large {\bf Instanton Moduli and Brane Creation }}

\vspace{10pt}
\auth

\vspace{10pt}

{\hoch{\dagger}\upenn}

\vspace{10pt}
{\hoch{\ddagger}\tamphys}

\vspace{40pt}

\underline{ABSTRACT}
\end{center}

     We obtain new intersecting 5-brane, string and pp-wave solutions
in the heterotic string on a torus and on a K3 manifold.  In the
former case the 5-brane is supported by Yang-Mills instantons, and in
the latter case both the 5-brane and the string are supported by the
instantons.  The instanton moduli are parameterised by the sizes and
locations of the instantons.  We exhibit two kinds of phase transition
in which, for suitable choices of the instanton moduli, a 5-brane
and/or a string can be created.  One kind of phase transition occurs
when the size of an instanton vanishes, while the other occurs when a
pair of Yang-Mills instantons coalesce.  We also study the associated
five-dimensional black holes and the implications of these phase
transitions for the black-hole entropy.  Specifically, we find that
the entropy of the three-charge black holes is zero when the
instantons are separated and of non-zero scale size, but becomes
non-zero (which can be counted miscrospically) after either of the
phase transitions.

{\vfill\leftline{}\vfill
\footnoterule
{\footnotesize \hoch{1} Research supported in part by DOE grant 
DE-FG02-95ER40893 \vskip -12pt} \vskip 14pt
{\footnotesize \hoch{2} Research supported in part by DOE grant
DE-AC02-76ER03071 \vskip -12pt} \vskip 14pt
{\footnotesize  \hoch{3} Research supported in part by DOE 
Grant DE-FG03-95ER40917.\vskip  -12pt}}

\pagebreak
\setcounter{page}{1}

\section{Introduction}

   The BPS $p$-branes of supergravity theories describe
non-perturbative states of the underlying string theory or M-theory.
In general, the $p$-brane solitons are not exact solutions in
supergravity, in the sense that delta-function singularities arise in
the field equations, implying that external source terms are needed.
These sources are in fact supplied by the associated fundamental
$p$-brane actions \cite{dkl}.  There are a few examples of $p$-brane
solitons in maximal supergravities where such source terms are
absent, notably the M5-brane \cite{guven} and the D3-brane
\cite{hs,dl}.

   In the heterotic string there is a different mechanism that can
give rise to regular brane-like solutions with no singular source
terms.  Due to the the Bianchi identity
\be
dF_\3 = \ft12 G_\2^a\wedge G_\2^a\ ,\label{bi}
\ee
one can construct a solitonic 5-brane that is supported by a
Yang-Mills instanton configuration living in the 4-dimensional space
transverse to the 5-brane worldvolume \cite{strom}.  This
configuration, unlike its 5-brane counterpart in the maximal $D=10$
supergravity, is a perfectly regular solution of the supergravity
equations of motion and is not supported by any external source
term.\footnote{One might argue that calling such a regular solution a
5-brane is somewhat inappropriate, and that it were better thought of
as an instanton solution which happens to have a Poincar\'e symmetry
in a six-dimensional submanifold.  We shall, however, follow the
traditional terminology and refer to it as the gauge 5-brane
\cite{strom}.}  The Bianchi identity (\ref{bi}) implies that the
5-brane charge is nothing but the total instanton number, providing a
natural quantisation of the 5-brane charge that lies outside, but is
consistent with, the usual Dirac quantisation condition.

    In this paper, we obtain a new solution describing the intersection
of a gauge 5-brane, a string and a pp-wave.  In other words, we show
that a string, with a wave propagating on its worldsheet, can lie on
the worldvolume of the instanton-supported 5-brane.  This
configuration is of particular interest since it reduces to a
three-charge black hole in $D=5$.  This means that we can study
thermodynamic quantities such as the entropy.

         The Yang-Mills instanton moduli are parameterised by the
sizes of the instantons and their locations.
The 5-brane charge, following from (\ref{bi}), is given by
\be
Q_m =\fft{1}{8\pi^2}\,  
\int dF_\3 = \fft1{16\pi^2}\,  \int G_\2^a\wedge G_\2^a = N\ ,
\ee
where the integration is over the entire 4-volume of the transverse
space and $N$ is the instanton number.  This charge is topological and
is therefore independent of the Yang-Mills instanton moduli.  This
leads to the interesting question as what happens if the size of a
instanton becomes zero, or if two instantons coalesce.  We show that
in each case there is a phase transition in which a fundamental
5-brane is created, while at the same time a gauge 5-brane is
destroyed.  Thus a gauge 5-brane turns into a fundamental 5-brane,
while keeping the total magnetic charge conserved.

       These two phase transitions have significant consequences for
the associated 3-charge black holes that arise after a dimensional
reduction to five dimensions.  The horizon has a curvature
singularity, and has zero area, when the instantons supporting the
gauge 5-brane are of non-vanishing size, and are non-coincident.
However, the horizon becomes regular (AdS $\times$ sphere), with
non-zero area, when either of the above phase transitions occurs.
This phenomenon supports the idea that a fundamental 5-brane is
created as a result of the phase transition, since the 3-charge black
hole in $D=5$ with non-zero horizon area can be interpreted, at the
microscopic level, by counting the states in such an intersecting
configuration \cite{sv}.

     We also study the phase transitions leading to brane creation in
the context of the heterotic string compacified on the K3 manifold.
In this case, both the string and 5-brane (in the ten-dimensional
picture) can be supported by Yang-Mills instantons; this corresponds
to gauge dyonic strings in $D=6$ \cite{dlpphase}.  Thus either of the
two kinds of phase transition discussed above will now lead to the
creation not only of fundamental 5-branes, but also fundamental
strings.  We obtain a new intersection with an additional superposed
pp-wave.  This gives rise, upon a further reduction to $D=5$, to
3-charge black holes with two instanton-supported charges and one
point charge, whose entropies become non-vanishing under either of the
two phase transitions.

        Supergravity on an anti-de Sitter spacetime background is
conjectured to be dual to an associated superconformal field theory on
its boundary \cite{mald}.  Thus the instanton phase transition can be
viewed as a transition from a supergravity theory to the superconformal
field theory.

\section{Heterotic string on torus}

          The low-energy effective action of the heterotic string is
$N=1$ supergravity in $D=10$, coupled to $E_8\times E_8$ Yang-Mills
matter fields.  We shall focus on an $SU(2)$ subgroup of $E_8\times
E_8$.  The Lagrangian for the bosonic sector is given by
\be
e^{-1}\, {\cal L}_{10} = R {*\oneone} -\ft12{*d\phi}\wedge {d\phi}
 -\ft12e^{-\phi}\, {*F_\3}\wedge F_\3 - \ft12 e^{-\fft12\phi}\, 
{*G^a_\2} \wedge G^a_\2\ ,\label{d10lag}
\ee
where the field $G^a_\2$ is the Yang-Mills field strength given by
\be
G_\2^\a = dB_\1^a + \ft12\epsilon^{abc}\, B_\1^b\wedge B_\1^c\ ,
\ee
and $F_\3$ is the three-form field strength, given by
\be
F_\3 = dA_\2 + \ft12 B^a_\1 \wedge dB_\1^a +\ft16 \epsilon^{abc}\,
B^a_\1 \wedge B^b_\1 \wedge B^c_\1\ .
\ee
It satisfies the Bianchi identity
\be
dF_\3 = \ft12 G^a_\2 \wedge G^\a_\2\ .\label{bianchi1}
\ee

\subsection{Instanton-supported intersections}

      The Lagrangian (\ref{d10lag}) admits a solution describing an 
intersection of a string, a 5-brane and a pp-wave, given by
\bea
ds_{10}^2 &=& H_e^{-3/4}\, H_m^{-1/4}\, (-W^{-1}\, dt^2 +
W\, (dz_1 + (W^{-1}-1)\, dt)^2)\nn\\
&& +H_e^{1/4}\, H_m^{-1/4}\, (dz_2^2 +\cdots + dz_5^2) +
 H_e^{1/4}\, H_m^{3/4}\, dy^i\, dy^i\ ,\nn\\
\phi&=&-\ft12 \log(H_e/H_m)\ ,\label{d10sol}\\
F_\3&=&  e^{\phi}\, {*(dt\wedge d^5z\wedge dH_m^{-1})} -
       dt\wedge dz_1\wedge dH_e^{-1}\ ,\nn
\eea
where the functions $H_e$, $H_m$ and $W$, associated with the string,
5-brane and pp-wave respectively, depend only on the four coordinates
$y^i$ of the transverse space, and satisfy the equations
\be
\square\, H_e=0\ ,\qquad \square \, W=0\ ,\qquad \square\, H_m =
-\ft14\, G^a_{ij}\, G^a_{ij}\ .\label{d10eqs}
\ee
Note that here $\square\equiv \del_i\, \del_i$ is the Laplacian in the
flat transverse metric $ds^2 =dy^i\, dy^i$, and the index contractions
in $G^a_{ij}\, G^a_{ij}$ are performed simply using the metric
$\delta_{ij}$ of the flat transverse space.  The $SU(2)$ Yang-Mills
fields $G_\2^a$ satisfy the self-duality equations
${*G^a_{ij}}=G^a_{ij}$ in the four-dimensional flat transverse space,
where $*$ denotes Hodge duality in this flat space.  Single-charge and
certain multi-charge $SU(2)$ instanton solutions are given in the
Appendix.

    For a single-center configuration, the solutions to equations 
(\ref{d10eqs}) can be taken to be:
\be
H_e = 1 + \fft{2Q_e}{r^2}\ ,\qquad
H_m= 1 +\fft{2(r^2 + 2a^2)}{(r^2 + a^2)^2}
\ ,\qquad
W=1 + \fft{2P}{r^2}\ ,\label{d10sols}
\ee
where we have made use of (\ref{1ymf2}) in order to solve the equation
of motion for $H_m$.  Note that the solution requires a fundamental
string as its source term, but does not require any fundamental
5-brane, since the 5-brane is supported by the Yang-Mills instanton,
which provides one unit of 5-brane charge.  This provides a
discretisation of the 5-brane charge that lies outside the Dirac
quantisation condition \cite{strom,dkl}.  This is possible due to
the Bianchi identity (\ref{bianchi1}).  Of course, if there were also
a fundamental 5-brane that could provide a delta-function source term,
then we could have an additional term $2\wtd Q_m/r^2$ in $H_m$, giving
\be
H_m=1 + \fft{2\wtd Q_m}{r^2} +  
\fft{2(r^2 + 2a^2)}{(r^2 + a^2)^2}\ .\qquad
\ee
(We shall discuss the normalisation of the magnetic charge below. Note
that a {\it unit} charge corresponds to a singularity of strength 2 in
the harmonic function.)

    For multi-centered configurations, we may take the solutions to
(\ref{d10eqs}) to be
\be
H_e = 1 + \sum_{\a'} 
\fft{2 Q_e^{\a'}}{|\vec y-\vec y{\, '}_{\!\!\a'}|^2}
\ ,\quad H_m = 1 + 2\psi\ ,\quad
W= 1 + \sum_{\a''} 
\fft{2P_{\a''}}{|\vec y - \vec y{\, ''}_{\!\!\a''}|^2}\ ,\label{mymsolh}
\ee
where $\psi$, given in (\ref{mymf2}), is associated with the
Yang-Mills instanton, discussed in detail in Appendix.  The primes on
the various indices and the locations of the singularities signify the
fact that the number, and locations, of the singularities for $H_e$,
$W$ and $f$ can all be different.

     To determine the approriate choice for the harmonic function $h$ in
(\ref{mymf2}), we must examine the behaviour of $\psi$ in the vicinity
of each singularity of the harmonic function $f$ in (\ref{fsol}).
Noting that $\square\,\log f = f^{-1}\, \square \, f -f^{-2}\, (\del_i
f)^2$, we see that near the singularity at $\vec y =\vec y_\a$, we
shall have $\psi = \ft14 \square\, \log f + h \sim - |\vec y-\vec
y_\a|^{-2} + h$, since $f\sim \lambda_\a\, |\vec y-\vec y_\a|^{-2}$
near this singularity.  We see from (\ref{mymsolh}) that in the
absence of any correction term from $h$, this would be of the form of
a singular point source with magnetic charge $(-1)$.  Therefore we
may exploit the freedom of adding an harmonic function $h = |\vec
y-\vec y_\a|^{-2}$ to $\psi$, in order to ensure that the only source
for the magnetic charge of the 5-brane at $\vec y = \vec y_\a$ is from
the Yang-Mills instanton.  Carrying out this procedure for each
singularity $\vec y_\a$, we see that $\psi$ in (\ref{mymsolh}) should
be chosen to be\footnote{In \cite{jnr}, the delta-function
singularities in $G^a_{ij}\, G^a_{ij}= -8\, \square\, \psi$ that result
from simply taking $\psi$ to be given by $\ft14\log f$ were eliminated
by excising small spheres around the singularities in $f$.  This could
be done there because $\psi$ itself had no direct physical
significance (and indeed it was not explicitly introduced in
\cite{jnr})).  In our case, however, $\psi$ itself appears when we
solve for $H_m$ in (\ref{d10eqs}) to obtain (\ref{d10sols}), and so we
must ensure that $\psi$ is free of singularities if we are to have a
solution that has only non-singular instanton source-terms for the
5-brane charge.  This procedure has the added advantage that
$G^a_{ij}\, G^a_{ij}= -8\, \square\, \psi$ is now an exactly correct
expression, with no delta-function singularities, and so the excision
of spheres performed in \cite{jnr} is no longer necessary.}
\be
\psi = \ft14 \square\, \log\Big( 1 + \sum_{\a=1}^N 
\fft{\lambda_\a}{|\vec y- \vec y_\a |^2} \Big) + \sum_{\a=1}^N 
\fft{1}{|\vec y -\vec y_\a|^2}\ .\label{psisol}
\ee

    In general, the total magnetic charge $Q_m$ is given by
\be
Q_m = \fft1{8\pi^2}\, \int_{S^3}\, F_\3\ .\label{qmag}
\ee
This can receive contributions both from non-singular
instanton-supported sources and from any singular sources
corresponding to the possible additional presence of point charges.
The instanton contributions can be calculated from the Bianchi
identity (\ref{bianchi1}), since the integral in (\ref{qmag}) can be
viewed as being over the sphere at infinity in the four-dimensional 
transverse space $V_4$, and hence we can write
\bea
Q_m &&= \fft1{8\pi^2}\, \int_{S^3}\, F_\3 = \fft1{8\pi^2}\, 
\int_{\rm V_4}\, dF_3\nn\\
&& = \fft1{16\pi^2}\, \int_{V_4} G_\2^a\wedge G_\2^2 \equiv N\ ,
\eea
where $N$ is the instanton number defined by the integral in
the second line.  If there are, in addition, point-charge singular
contributions, then these can be calculated from the expression for 
$F_\3$ in (\ref{d10sol}).  The first term gives the magnetic
contribution $F_\3 = -\ft16 \del_i\, H_m\, \ep_{ijk\ell}\, dy^j\wedge
dy^k\wedge dy^\ell$, and hence a contribution to $dF_\3$ of 
$dF_\3= -\square\, H_m\, d^4y$.  This implies that there will be 
a contribution to the magnetic charge 
\be
Q_m = \fft{1}{8\pi^2}\, \int_{S^3}  F_\3 = -\fft1{8\pi^2}\,
\int_{V_4}\, \square H_m\ .
\ee
Thus a term in  $H_m$ of the form $2k\, |\vec y - \vec y_\a|^{-2}$ will 
contribute a magnetic charge 
\be
Q_m = -\fft1{4\pi^2}\, \int_{S_3}\, \del_i\Big(\fft{k}{|\vec y-\vec
y_\a|^2} \Big)\, d\Sigma_i = k\ .
\ee

    Putting this all together, we see that a solution with $N$
instantons with scales $\lambda_\a$ centered at the points $\vec y_\a$,
and $N'$ point magnetic charges $\wtd Q_m^{\td\a}$, centered at the
points $\vec y_{\td\a}$, will be described in terms of a function
$H_m$ in (\ref{mymsolh}) with
\be
\psi = \Big[ \ft14 \square\, \log\Big( 1 + \sum_{\a=1}^N 
\fft{\lambda_\a}{|\vec y- \vec y_\a |^2} \Big) + \sum_{\a=1}^N 
\fft{1}{|\vec y -\vec y_\a|^2}\Big] + \sum_{\td\a=1}^{N'}\, 
\fft{\wtd Q_m^{\td \a}}{|\vec y-\vec y_{\td\a}|^2}\ .
\ee
The term enclosed in square brackets is the non-singular contribution
from the instantons, and the final term is the singular contribution
of the point charges. The total magnetic 5-brane charge will be
\be
Q_m = N + \sum_{\td\a=1}^{N'} \wtd Q_m^{\td \a}\ .
\ee

\subsection{Brane creation}

   We have seen that the moduli space of the instantons in the
solutions we are discussing is parameterised by the sizes of the
instantons $\lambda_\a$ and their positions $\vec y_\a$.  Two types
of phase transitions can arise when one adjusts these modulus
parameters.  The first type is associated with the sizes of the
instantons.  If the scale-size of an instanton located at $\vec y=\vec
y_\a$ is taken to zero, there is a point singularity left at $\vec
y_\a$.  To see this explicitly, we note that in the vicinity of the
instanton location $\vec y_\a$, the function $H_m$ defined by
(\ref{mymsolh}) and (\ref{psisol}) becomes
\be
H_m=1 + \fft{2(r^2 + 2a^2)}{(r^2 + a^2)^2}\phantom{xxx} 
\mapright{a\rightarrow 0}
\phantom{xxx} 1 + \fft2{r^2}\ ,\label{vanishingsize}
\ee
where $\vec r = \vec y -\vec y_\a$, and $a=\sqrt\lambda_\a$ is the
scale-size of the instanton.  In other words, the function
$H_m$ becomes a harmonic function, associated with a point singularity
in the transverse space, when the instanton size vanishes. This point charge,
unlike the case of the non-singular instanton, has a delta-function
singularity, implying the need for a source term outside the $N=1$, $D=10$
supergravity.  This external source is in fact provided by introducing
a fundamental 5-brane action.  Thus we see that a fundamental 5-brane is
created when the instanton size is taken to zero.  In this phase
transition, the total magnetic charge measured by $\int F_\3$ is
conserved.

   Another kind of phase transition occurs if two of the instanton
centers are allowed to become coincident.  Suppose, for example, that
we take $\vec y_\a = \vec y_\b$ for two specific instanton centers
$\vec y_\a$ and $\vec y_\b$.  In the function $f= 1 + \sum_\a
\lambda_\a\, |\vec y-\vec y_\a|^{-2}$, the effect is merely to
coalesce a two-instanton configuration with instantons of size
$a^2=\lambda_\a$ and ${a'}^2=\lambda_\beta$ into a {\it one}-instanton
configuration of size ${a''}^2 = \lambda_\a + \lambda_\beta$.
However, the harmonic function $h$ will now have a term $2|\vec y-\vec
y_\a|^{-2}$, whose strength is twice the value that is needed for
cancelling out the singularity in $\psi$ at $\vec y = \vec y_\a$.
Thus, there is one unit of point charge (in the transverse space) left
over.  The upshot of this is that when two instanton centers are
brought into coincidence, a configuration that previously described a
non-singular gauge 5-brane with instanton number 2 undergoes a phase
transition to a configuration describing two superposed 5-branes
supported by one non-singular instanton charge and one singular
point-magnetic-charge, which is nothing but the fundamental 5-brane
charge.  In particular, as must be, the net magnetic charge is
conserved.

    As an example, consider a 2-instanton solution, where the two
instanton centers are initially located at $\vec y_1$ and $\vec y_2$.
The function $\psi$ will be given by
\be
\psi = \ft14\square\,\log\Big(1 + \fft{\lambda_1}{|\vec y-\vec y_1|^2} 
+  \fft{\lambda_2}{|\vec y-\vec y_2|^2}\Big) + \fft1{|\vec y-\vec
y_1|^2} +  \fft1{|\vec y-\vec y_2|^2}\ .
\ee
After allowing the instanton centers to coalesce, say at $\vec y=\vec
y_1$, the function $\psi$ becomes
\be
\psi = \Big[\ft14\square\,\log\Big(1 + 
\fft{\lambda_1+\lambda_2}{|\vec y-\vec y_1|^2} \Big) + \fft1{|\vec y-\vec
y_1|^2}\Big] +  \fft1{|\vec y-\vec y_1|^2}\ ,
\ee
where the term enclosed in square brackets is the non-singular
contribution to the function $H_m$ in (\ref{d10sols}) coming from the
remaining instanton-supported charge, while the final term describes
the singular contribution to $H_m$ coming from a unit point-charge
located at $\vec y_1$. 

       In both of the phase transitions that lead to the creation of
fundamental 5-branes, the fundamental 5-brane charge that is generated
is quantised and is equal to the decrease in the Yang-Mills instanton
number.  The 5-branes that are created all contribute positively to
the total mass.

    One could have asked the question purely in the framework of
four-dimensional Yang-Mills theory as to what happens when two
instanton centers in a multi-instanton solution are allowed to
coalesce.  However, unlike the situation that we have been discussing
here, the question in four-dimensional Yang-Mills theory is an
entirely non-dynamical one, in the sense that there is no external
``time'' coordinate and, thus, no possibility of a ``slow motion''
confluence of the instanton centers.  Thus four-dimensional Yang-Mills
theory does not really in itself demand that one give a precise
interpretation to the question of what happens if two instanton
centers coalesce.  In our case, however, where the instantons reside
in a four-dimensional space transverse to the 5-branes, it does make
sense to envisage a slow-motion approximation in which the locations
of the instanton centers vary as a function of time.  Thus it is
important in this context that one should be able to give a sensible
interpretation, of the kind that we have supplied, to the question of
what happens when two instanton centers coalesce.

\subsection{$D=5$ black hole and its entropy}

        The intersection solution (\ref{d10sol}) is invariant under
translational symmetry of the coordinates $\{z_1, z_2, \ldots, z_5\}$.
It follows that we can dimensionally reduce the solution on the
five-torus $T^5$ associated with these coordinates, giving rise to a
three-charge black hole in $D=5$.  This torus reduction can also be
consistently performed on the Lagrangian (\ref{d10lag}).  The relevant
part of the five-dimensional Lagrangian of the associated three-charge
black hole is given by
\bea
e^{-1}\, {\cal L}_5&=& R\, {*\oneone} -\ft12 {*d\vec \phi}\wedge
d\vec\phi -\ft12 e^{\vec a\cdot \vec \phi}\, {* F_\3}\wedge F_\3 -
\ft12 e^{\vec d\cdot \vec \phi}\, {*F_\2}\wedge F_\2\nn\\
&& -\ft12 e^{\vec b \cdot \vec \phi}\, {*\cF_\2}\wedge \cF_\2 -
\ft12  e^{\vec c\cdot\vec\phi}\, {*G_\2^a}\wedge G_\2^a\ ,\label{d5lag1}
\eea
where $\cF_\2 =d\cA_\1$ is the Kaluza-Klein two-form field strength.
The dilaton vectors $\vec a$, $\vec b$, $\vec c$ and $\vec d$ in
(\ref{d5lag1}) satisfy the following product rules
\bea
&&\vec a\cdot \vec a = \vec b \cdot \vec b = 
\vec d\cdot \vec d = 4 \vec c \cdot \vec c = \ft83\ ,\qquad
\vec a \cdot \vec b = \vec a\cdot \vec d=
\vec a \cdot \vec c = \ft43\ ,\nn\\
&& \vec b\cdot \vec c=\vec d \cdot \vec c = \ft23 \ ,\qquad
\vec b\cdot \vec d =-\ft43\ .
\eea
We can realise these by the two-component
vectors
\be
\vec a=(-\sqrt2, \sqrt{\ft23})\ ,\qquad
\vec b=(0, \sqrt{\ft83})\ ,\qquad
\vec c=(-\sqrt{\ft12}, \sqrt{\ft16})\ ,\qquad
\vec d=(-\sqrt2, -\sqrt{\ft23})\ .\label{abcd}
\ee

        The three-charge five-dimensional black hole, which is the
dimensional reduction of (\ref{d10sol}) and a solution to
(\ref{d5lag1}),  is given by
\bea
ds_5^2 &=& -(H_e\, H_m\, W)^{-2/3}\, dt^2 + (H_e\, H_m\, W)^{1/3}\, 
(dr^2 + r^2\, d\Omega_3^2)\ ,\nn\\
\vec \phi &=& - \ft12 \vec a\, \log H_m +
\ft12 \vec b\, \log W + \ft12 \vec d \, \log H_e\ ,\label{d5sol1}\\
\cF_\2 &=& dt\wedge dW^{-1}\ ,\qquad
F_\2 = dt\wedge dH_e^{-1}\ ,\qquad
F_\3 = e^{-\vec a\cdot \vec\phi}\, {*(dt\wedge dH_m^{-1})}\ .\nn
\eea
For convenience, we have assumed here, as we did for the previous
gauge 5-brane solution, that the asymptotic values of the dilatons
vanish; $\vec \phi_0=0$.  Here, for simplicity, we consider only the
isotropic black hole, where all the charges are located at the origin.
In this case, there is only a single instanton, contributing one unit
of the charge associated with $F_\3$.  The metric in (\ref{d5sol1})
has an horizon at $r=0$.  For any non-vanishing size $a$ of the
instanton, the metric (\ref{d5lag1}) is singular at the horizon, which
has vanishing area.  It follows that the entropy is exactly zero.  On
other hand, when the instanton size is zero, the instanton is replaced
by a point charge in the transverse space.  In this case, the horizon
becomes regular and has a non-zero area.  Thus the entropy undergoes a
phase transitition as the scale-size of the instanton vanishes:
\be
S=\left\{ \begin{array}{r@{\qquad : \qquad}l}
         0 &  a >0 \\
         \ft14 A_{\rm horizon} = \pi^2\, \sqrt{2Q_e\, P} & a =0
          \end{array} \right. 
\ee

        An analogous phenomenon occurs if $N+1$ instantons coalesce.
The entropy, which is initially zero, becomes non-vanishing and is
given by 
\be
S=\pi^2 \sqrt{2N Q_e\, P}\ .
\ee
This non-vanishing of the area of the horizon in either of the two
kinds of phase transition supports the earlier proposal that 5-branes
are created, since the entropies of these black-hole configurations
can be independently evaluated in terms of a miscroscopic counting of
string states propagating on the intersecting D1-D5 brane system
\cite{sv}.  One might envisage that although the black hole entropy,
which is equal to one quarter of the area of the black hole event
horizon, vanishes when the Yang-Mills instanton size is non-zero or
the instantons are seperated, it is possible that the total entropy,
which is the sum of the black hole entropy and the entropy of the
Yang-Mills excitations, may be conserved in the phase transition.  It
is worth mentioning that the dilaton behavior is quite different
before and after the phase transition.  Before the phase transition,
the dilatonic scalars diverge on the horizon, with the consequence that
the classical black-hole solution is not reliable for extracting
information about physical quantities such as the entropy.  After the phase
transition, in which a fundamental 5-brane is created, the dilatons
are stablised on the horizon, and consequently the non-vanishing
entropy can be evaluated by independent microscopic methods.

    We have seen that the horizon has a curvature singularity when the
instantons are of non-zero size and are separated but that, under
either of the phase transitions, the horizon becomes regular once the
point-source limit is reached.  In $D=5$, the near-horizon structure
after the phase transition is AdS$_2\times S^3$.  From the
ten-dimensional point of view, it is AdS$_3\times S^3 \times T^4$.
The AdS$_3$ is also known as the extremal BTZ black hole \cite{btz},
which is a special case of the generalised Kaigorodov metric
\cite{clp,kaig}.  Supergravity on this AdS$_3$ background is
conjectured to be dual to a two-dimensional superconformal field
theory on the boundary of the AdS$_3$ \cite{mald}.  Thus the instanton
phase transition can be viewed as a transition from supergravity
theory to a two-dimensional conformal field theory.

      Another physical quantity that undergoes a phase transition is
the absorption rate for massless scalar waves.  For the case when the
instanton size $a$ is non-zero, the near-horizon structure of the
black hole is dominated by the electric string and wave charges and
its low energy absorption cross-section is proportional to the
frequency of the wave \cite{clpt}.  On the other hand, when the
instanton scale size $a$ goes to zero, the absorption cross-section
approaches the area of the horizon in the low-frequency limit.  To
summarise, we have
\be
\sigma \sim  \left\{ \begin{array}{r@{\qquad : \qquad}l}
         2\pi^2\,  (Q_e\, P)\, \omega  &  a >0 \\
         A_{\rm horizon} & a =0
          \end{array} \right. 
\ee

\section{Heterotic string on K3}

       In the previous section, we obtained intersections of a string,
a 5-brane and a pp-wave in the heterotic string, where the 5-brane
carries magnetic charge supported by a Yang-Mills instanton or
multi-instanton configuration.  When dimensionally reduced on a torus
to $D=6$, the intersection becomes that of a dyonic string with a
pp-wave, where the magnetic charge of the string is supported by the
Yang-Mills instanton, while the electric charge is associated with
singular sources.  Thus, in this case, the electric and the magnetic
strings play very different r\^oles.  In this section, we shall
consider the heterotic string compactified on K3 rather than a
4-torus, in which case not only the magnetic strings, but also the
electric strings, can be supported by Yang-Mills instantons.

\subsection{$N=1$ supergravity in $D=6$}

    The heterotic string admits a compactification to $D=6$ in which
the internal four-dimensional manifold is taken to be K3.  Various
different six-dimensional theories can be obtained, with different
Yang-Mills gauge groups, depending upon precisely how the
$SU(2)$-valued spin connection of the Ricci-flat K\"ahler K3 is
embedded in the $E_8\times E_8$ or $SO(32)$ gauge group of the
ten-dimensional theory \cite{sag1}.  There will also be quantum
corrections to the six-dimensional effective action, whose 1-loop
structures can be determined by general arguments based on the
necessary anomaly-freedom of the theory.  The resulting
six-dimensional theories are described by $N=1$ supergravity, coupled
to an $N=1$ hypermultiplet and a Yang-Mills multiplet.  The bosonic
sectors comprise the metric $g_{\mu\nu}$, a dilaton $\phi$, a 3-form
field strength $F_\3$, and the Yang-Mills fields $G^a_\2$.  The
self-dual part of the 3-form field belongs to the gravity multiplet,
while the anti-self-dual part and the dilaton belong to the
hypermultiplet.  The field equations, including the 1-loop terms, take
the form \cite{sag1}
\bea
R_{\mu\nu} &=& \ft12 \del_\mu\phi\, \del_\nu\phi + \ft14
e^{-2\a\phi}\, [F^2_{\mu\nu} - \ft16 F_\3^2\, g_{\mu\nu}]\nn\\
&& + \ft14 (v\,
e^{-\a\phi} + \td v\, e^{\a\phi})\, [ (G^a)^2_{\mu\nu} - \ft18
(G^a_\2)^2\, g_{\mu\nu}]\ ,\nn\\
d{*d\phi} &=& \a\, e^{-2\a\phi}\, {*F_\3}\wedge F_\3 + \ft12\a\, (v\,
e^{-\a\phi} - \td v\, e^{\a\phi})\, {*G_\2^a}\wedge G_\2^a\ ,\nn\\
d(e^{-2\a\phi}\,{*F_\3}) &=& \ft12 v\, G_\2^a\wedge G_\2^a\ ,\label{sag}\\
D[(v\,e^{-\a\phi} + \td v\, e^{\a\phi})\, {*G_\2^a}] &=& v\,
e^{-2\a\phi}\, {*F_\3}\wedge G_\2^a + \td v\, F_\3\wedge G_\2^a\ ,\nn
\eea
where $\a=1/\sqrt2$. Here, $D$ denotes the Yang-Mills-covariant
exterior derivative, defined by
\be
D X^a = d X^a -\ep_{abc}\, X^b\wedge B_\1^c\ ,
\ee
where, as previously, we restrict attention to an $SU(2)$ subgroup of
the Yang-Mills gauge group.  The constants $v$ and $\td v$ are
rational numbers characteristic of the embedding of the $SU(2)$ holonomy group
of K3 in the original $E_8\times E_8$ or $SO(32)$ Yang-Mills gauge
group in $D=10$.  The terms associated with $\td v$ come from 1-loop
corrections.  The field strengths are given in terms of potentials as
follows:
\bea
F_\3 &=& d A_\2 + \ft12 v\, \omega\ ,\nn\\
G_\2^a &=& dB_\1^a + \ft12 \ep_{abc}\, B_\1^b\wedge B_\1^c\ .
\eea
Here, $\omega$ is given by
\be
\omega = B_\1^a\wedge dB_\1^a + \ft13 \ep_{abc}\, B_\1^a\wedge
B_\1^b\wedge B_\1^c\ ,
\ee
and by construction it satisfies $d\omega = G_\2^a\wedge G_\2^a$.

   The field equations (\ref{sag}) cannot be obtained from any
Lagrangian.  However, there is a closely-related system of field
equations which, in particular, admit the same set of solutions that
we wish to consider, which {\it can} be derived from a Lagrangian.
If we consider the Lagrangian
\be
{\cal L}_6 = R\, {*\oneone} - \ft12 {*d\phi}\wedge d\phi - \ft12
e^{-2\a\phi}\, {*F_\3}\wedge F_\3 - \ft12 (v\, e^{-\a\phi} + \td v\,
e^{\a\phi})\, {*G_\2^a}\wedge G_\2^a + \ft12 \td v\, G_\2^a\wedge
G_\2^a\wedge A_\2\ ,\label{nonsaglag}
\ee
it is easily seen that it correctly produces all except one of the
equations of motion given in (\ref{sag}).  The exception is the
Yang-Mills equation, which turns out to be
\be
 D[(v\,e^{-\a\phi} + \td v\, e^{\a\phi})\, {*G_\2^a}] = v\,
e^{-2\a\phi}\, {*F_\3}\wedge G_\2^a + \td v\, dA_\2\wedge G_\2^a
-\ft14 v\, \td v\, G_\2^a\wedge G_\2^a\wedge B_\1^a\ ,\label{nonsag}
\ee
rather than the corresponding equation in (\ref{sag}).  
   
    The discrepancy between the Yang-Mills equations in (\ref{sag}) and
(\ref{nonsag}) is a term of the form
\bea
&&(F_\3 -dA_\2) \wedge G_\2^a  +\ft14 v\, G_\2^b\wedge G_\2^b\wedge B_\1^a
\nn\\
&&= \ft12 v\, \Big[(B_\1^b\wedge dB_\1^b + \ft13 \ep_{bcd}\,
B_\1^b\wedge B_\1^c\wedge B_\1^d)\wedge G_\2^a +\ft12  G_\2^b\wedge
G_\2^b\wedge B_\1^a \Big]\ .
\eea
It is therefore evident, since this involves only the Yang-Mills
fields, that if we consider instanton solutions where $B_\1^a$ is
non-vanishing only in the four-dimensional transverse space, then this
5-form will vanish.  Thus for such configurations, the solutions of
(\ref{sag}) and those following from (\ref{nonsaglag}) will coincide.
Note that the theory admits two different types of global limit
\cite{dllp}.  In one of the limits, the resulting flat-space theory
admits a tensionless string as a solution \cite{dlpphase}. The other
is a further specialisation of the flat-space theory and had also
been obtained in \cite{bss}.

\subsection{Gauge dyonic strings with pp-wave, and $D=5$
black hole}

        The equations of motion (\ref{sag}) admit solutions describing
the intersection of a dyonic string with a
pp-wave, given by
\bea
ds_6^2&=&(H_e\, H_m)^{-1/2}(-W^{-1}\, dt^2 + W(dz+ (W^{-1} -1) dt)^2
       + (H_e\, H_m)^{1/2}\, dy^i\, dy^i\ ,\nn\\
\phi&=&\fft1{\sqrt2}\log(H_m/H_e)\ ,\nn\\
F_\3&=& e^{-\sqrt2\phi}\, {*(dt\wedge dz\wedge dH^{-1}_m)} -
                dt\wedge dz\wedge dH_e^{-1}\ .\label{d6sol}
\eea
Here, $H_e$, $H_m$ and $W$ satisfy
\be
\square\, H_e = -\ft14 \td v\, G_{ij}^a\, G_{ij}^a\ ,
\qquad \square\, H_m =-\ft14 v\, G_{ij}^a\, G_{ij}^a\ ,\qquad
\square\, W=0\ .
\ee
Thus we have
\be
H_e = 1 + 2\td v\, \psi\ ,\qquad
H_m = 1 + 2v\, \psi\ ,\qquad
W= 1 + \sum_{\a'} 
\fft{2P_{\a'}}{|\vec y - \vec y{\, '}_{\a'}|^2}\ ,\label{mymsolh2}
\ee
where for multi-instantons, $\psi$ is given by (\ref{psisol}).  (The
solution with no pp-wave was obtained in \cite{dlpphase}.)  Thus we
see that when the size of an instanton vanishes, or when two
instantons coalesce, there is a creation not only of a magnetic
string, coming from the dimensional reduction of the 5-brane in
$D=10$, but also of an electric string.

        The dyonic string solution (\ref{d6sol}) can be dimensionally
reduced on the $z$ coordinate, giving rise to a $D=5$ three-charge
black hole.  The form of the solution is the same as given in
(\ref{d5sol1}), except that now the functions $H_e$, $H_m$ are given
by (\ref{mymsolh2}) instead of (\ref{mymsolh}).  Equations of motion that
describe these black holes can be derived from the five-dimensional 
Lagrangian
\bea {\cal L}_5&=& R\, {*\oneone} -\ft12 {*d\vec \phi}\wedge d\vec\phi
-\ft12 e^{\vec a\cdot \vec \phi}\, {* F_\3}\wedge F_\3 - \ft12 e^{\vec
d\cdot \vec \phi}\, {*F_\2}\wedge F_\2\nn\\ 
&& -\ft12 e^{\vec b \cdot
\vec \phi}\, {*\cF_\2}\wedge \cF_\2 - \ft12 v\, e^{\vec
c\cdot\vec\phi}\, {*G_\2^a}\wedge G_\2^a -\ft12 \td v\, e^{\vec {\td
c}\cdot\vec\phi}\, {*G_\2^a}\wedge G_\2^a\nn\\ 
&& +\ft12 \td v\,
A_\1\wedge G_\2^a\wedge G_\2^a \ ,\label{d5lag2} 
\eea 
where the dilaton vectors $\vec a$, $\vec b$, $\vec c$ and $\vec d$
are given by (\ref{abcd}), and $\vec {\td c} = (1/\sqrt2, 1/\sqrt6)$.
This is obtained by dimensional reduction of the $D=6$ Lagrangian
(\ref{nonsaglag}).  Again, this produces equations of motion which do
not coincide precisely with those of the dimensionally-reduced string
(which cannot themeselves be derived from a Lagrangian).  However, the
discrepancies between the string equations of motion and those
following from (\ref{d5lag2}) are terms which vanish for the
configurations we are considering.

        The discussion of the entropy of the 3-charge black hole is
analogous to the previous case.  When the instanton size is non-zero,
the entropy vanishes; when the size becomes zero, the area of the
horizon becomes non-zero and hence the entropy is non-vanishing.  The
singular horizon becomes regular and, in terms of the ten-dimensional
point of view, the near horizon structure is AdS$_3\times
S^3\times$K3.  Before the phase transition, the metric has an 
{\sl almost} naked singularity, which can be reached in a
logarithmically-divergent time by a null geodesic.  
A closely-related feature is that the absorption cross-section for 
scalar waves vanishes below a certain frequency
threshold \cite{clpt}.  After the phase
transition, the metric becomes regular and at low frequencies the absorption 
cross-section is approximately equal to non-vanishing area of the
horizon.  Prior to any phase transition, 
there is one significant difference between this solution and the one 
obtained from the heterotic string on
torus, which we discussed previously.  Here, owing to the fact that both 
the string and the 5-brane are
supported by Yang-Mills instantons, it follows that the six-dimensional
dilaton remains finite as $r$ tends to zero.  Thus the six-dimensional dyonic
string with a pp-wave has a regular horizon both before and after the phase
transition.  This should be contrasted with the previous 5-brane example,
where the associated intersection has a regular horizon only after the phase
transition occurs.

\section{Conclusions}

    In this paper, we have studied certain extremal $p$-brane
configurations in which one or more of the charges are supplied by
Yang-Mills instantons in a four-dimensional transverse space.
Previously known examples were the gauge 5-brane in the
ten-dimensional heterotic theory \cite{strom} and the gauge dyonic
string in the theory in six dimensions obtained by compactifying the
heterotic string on K3 \cite{dlpphase}.  If the gauge 5-brane is
compactified on $T^5$, or the gauge dyonic string is compactified on
$S^1$, one obtains in either case a five-dimensional black hole.  The
former gives a 1-charge magnetic black hole, while the latter gives a
2-charge dyonic black hole.  In both cases, the charges are ``smeared
out'' by the Yang-Mills instanton construction.

   It is of interest to study configurations that correspond to
3-charge black holes in five dimensions, since then one has the
possibility of having a non-zero entropy even for extremal
configurations.  For this reason, we constructed generalisations of
the previously-known gauge solutions, namely a gauge 5-brane
intersecting with a string and a pp-wave in $D=10$ and a gauge dyonic
string intersecting with a pp-wave in $D=6$.  These give rise to
five-dimensional 3-charge black holes with one smeared charge or
two smeared charges respectively, with the remainder being standard
point-source charges.

   We showed that, as long as the Yang-Mills instantons are
non-degenerate, the entropies of the 3-charge black holes vanish.
Indeed, from this point of view, the smeared charges coming from the
instantons contribute little to the horizon structures and the black
holes are more like those with the correspondingly fewer number of
``genuine'' point charges.  However, we also showed that, if certain
singular limits of the instanton configurations are taken, the
resulting black holes undergo phase transitions in which they acquire
the non-vanishing entropy associated with the usual 3-charge black
holes.

   We exhibited two different kinds of degenerate limits for the
Yang-Mills instanton configurations, each of which leads to such phase
transitions.  One of these is the situation where the scale-size of an
instanton goes to zero, leading to the appearance of a single unit of
point charge at the location of the associated instanton center.
Another, perhaps more surprising, degenerate limit occurs if two
previously-separated instantons come into coincidence.  This leads to
a configuration with a superposed instanton and a unit point charge at
the coincidence point.  In either of these cases, the emergence of the
point charge in the transverse space in the singular limit gives rise
to the phase transition.  This singularity is nothing but the
fundamental 5-brane or string charge.

\section*{Appendix}

\appendix

\section{$SU(2)$ Yang-Mills instantons}

    The solutions that we consider in this paper all involve the use
of an $SU(2)$ Yang-Mills instanton in the four-dimensional transverse
space.  The simplest such solution is the BPST single instanton, which
is spherically symmetric.  This is most elegantly described by writing
the metric on the flat transverse space in terms of hyperspherical
polar coordinates, as
\be
ds^2 = dr^2 + \ft14 r^2\, (\sigma_1^2 + \sigma_2^2 + \sigma_3^2)\ ,
\label{hspolar}
\ee
where the $\sigma_a$ are the three left-invariant 1-forms on the
3-sphere, satisfying the equation $d\sigma_a = -\ft12 \ep_{abc}\,
\sigma_b\wedge \sigma_c$.  The instanton is obtained by making the
ansatz
\be
B_\1^a = h\, \sigma_a\ ,\label{ymans}
\ee
where $h$ is a function only of $r$.  A simple symmetrical ansatz of
this type is possible because we are considering a Yang-Mills instanton
with $SU(2)$ gauge group, which coincides with the left-acting
symmetry group of the 3-sphere.  It is elementary to calculate the
Yang-Mills field strengths $G_\2^a$ for the ansatz (\ref{ymans}) and
then to show that the self-duality equations are satisfied if $r\, h'
= 2h\, (h-1)$.  The general solution of this equation is 
\be
h = \fft{a^2}{a^2+r^2}\ ,
\ee
where $a$ is an arbitrary constant which sets the scale-size of the
instanton.  The Yang-Mills field strength is therefore given by
\be
G_\2^a = -\fft{4 a^2}{(a^2+r^2)^2}\, (e^0\wedge e^a + \ft12
\ep_{abc}\, e^b\wedge e^c)\ ,\label{1ymsol}
\ee
where $e^0=dr$ and $e^a=\ft12 r\, \sigma_a$ is a vielbein basis for 
(\ref{hspolar}).  Note that $G_\2^a$ is manifestly self dual.  One
easily verifies from (\ref{1ymsol}) that $G^a_{ij}\, G^a_{ij} = 
192 a^4\, (a^2+r^2)^{-4}$ and, hence, that 
\be
G^a_{ij}\, G^a_{ij} = -8\, \square \psi\ ,\qquad {\rm where}\qquad
\psi = \fft{r^2+ 2a^2}{(r^2+a^2)^2}\ ,\label{1ymf2}
\ee
where $\square$ here denotes the scalar Laplacian in the
four-dimensional flat transverse-space metric (\ref{hspolar}).
Note that the local solution for $\psi$ is ambiguous up to the
addition of a harmonic term $k/r^2$, and we have resolved this
ambiguity by choosing $k$ so that $\psi$ has no singularity at $r=0$.

     The general multi-instanton solutions are most completely
described by the ADHM construction \cite{adhm}.  Sub-classes of
solution are describable using more elementary methods
\cite{thooft,jnr}.  For this purpose, it is convenient to write the
metric on the four-dimensional transverse space in Cartesian
coordinates $y^i$, for $i=0,1,2,3$, as $ds^2=dy^i\, dy^i$.  Let us
define the anti-self-dual 't Hooft tensors $\eta^a_{ij}$, which are
antisymmetric and anti-self-dual in $ij$.  Thus
\be
\eta^a = \ft12 \eta^a_{ij}\, dy^i\wedge dy^j = 
-dy^0\wedge dy^a + \ft12\ep_{abc}\, dy^b\wedge dy^c\ .
\ee
In other words, $\eta^a_{0b}= -\delta^a_b$, $\eta^a_{b0} = \delta^a_b$
and $\eta^a_{bc}=\ep_{abc}$.  The ansatz for the Yang-Mills potentials
is
\be
B_\1^a = -\eta^a_{ij}\, \del_i \tilde f\, dy^j\ .
\ee
After a little algebra, one finds that self-duality ${*G_\2^a}=G_\2^a$
implies the equation $\square \td f + \del_i \td f\, \del_i \td f=0$,
which is solved by taking $\td f=\log f$, where $f$ satisfies $\square
f=0$.  Thus we have instanton solutions with
\be
f = \ep + 
\sum_{\a=1}^N \fft{\lambda_\a}{|\vec y-\vec y_\a|^2}\ ,\label{fsol}
\ee
where $\ep$ is a constant that can be taken to be either 1 or 0, and
$\lambda_\a$ and $\vec y_\a$ are constant strengths and positions for
the singularities in $f$.  When $\ep=1$, they have rather direct
interpretations as scale sizes and positions for $N$ separated
Yang-Mills instantons \cite{thooft}.  When $\ep=0$, the interpretation
is more subtle and (\ref{fsol}) then actually describes an
$(N-1)$-instanton solution, with scale sizes and locations that are
rather complicated functions of the $\lambda_\a$ and $\vec y_\a$
parameters.  

    After further algebra, one can show that, for these multi-instanton
solutions, we have
\be
G^a_{ij}\, G^a_{ij} = -2\, \square\, \square \, \log f\ .
\ee
This means that we may write $G^a_{ij}\, G^a_{ij}$ as
\be
G^a_{ij}\, G^a_{ij} = -8\, \square\psi\ ,\qquad {\rm where}\qquad \psi
=\ft14 \square \log f + h\ ,\label{mymf2}
\ee
$f$ is the harmonic function given in (\ref{fsol}) and $h$ is an
arbitrary harmonic function.  As in the single-instanton example
discussed above, we may exploit the freedom to add such an harmonic
function in order to ensure that $\psi$ itself is non-singular at the
locations of the instantons.  This is discussed in section 2. We
consider, for convenience, the case where $\ep=1$ in (\ref{fsol}),
since then the parameters $\lambda_\a$ and $\vec y_\a$ have clearer
interpretations.

   To see how the parameters may be interpreted, consider the special
case $N=1$ in (\ref{fsol}), with $\ep=1$.  Without loss of generality,
we may take $\vec y_\a=0$ and $\lambda_\a=\lambda$, so that
\be
f = 1 + \fft{\lambda}{r^2}\ ,
\ee
where $r=|\vec y|$.  Evaluating $\psi$ as given in (\ref{mymf2}), with
$h$ chosen to be $1/r^2$, we obtain $\psi = (2\lambda + r^2)(\lambda +
r^2)^{-2}$.  Comparing with (\ref{1ymf2}), we see that the $N=1$ solution 
has precisely the interpretation of a single Yang-Mills instanton of
size $a=\sqrt{\lambda}$, located at $\vec y=0$.  The general
$N$-instanton solution (\ref{fsol}) with $\ep=1$ describes instantons
of size $\sqrt{\lambda_\a}$ centered on locations $\vec y_\a$.

\end{document}